\input ./psfig.tex
\documentstyle[]{mn}
\def\mch{M$\rm^{c}$Hardy~}
\def\etal{et al.~\rm}

\def\erg{{\rm\thinspace erg}}
\def\cm{{\rm\thinspace cm}}

\def\s{{\rm\thinspace s}}
\def\ergps{\hbox{$\erg\s^{-1}\,$}}
\def\ecs{\hbox{$\erg\cm^{-2}\s^{-1}\,$}}

\def\h50{\hbox{h$_{50}$}}
\def\m12{\hbox{$\Delta$m$_{12}$}}
\def\lesssim{\mathrel{\hbox{\rlap{\hbox{\lower4pt\hbox{$\sim$}}}\hbox{$<$}}}}
\def\gtrsim{\mathrel{\hbox{\rlap{\hbox{\lower4pt\hbox{$\sim$}}}\hbox{$>$}}}}
\def\logn{log $N$-log $S$ relation~}

\title[X-ray evolution of galaxy groups]
{The detection and X-ray evolution of galaxy groups at high redshift}

\author[Jones, L.R. et al.]
{
L. R. Jones,$^{1,5}$ I. \mch$^2$, A. Newsam$^3$ and K. Mason$^4$\\
$^1$School of Physics \& Astronomy, University of Birmingham,
Birmingham B15 2TT.\\
$^2$Department of Physics and Astronomy, The University, Southampton SO17 1BJ\\
$^3$Astrophysics Research Institute, Liverpool
John
Moores University, Twelve Quays House, Egerton Wharf, \\
Birkenhead CH41 1LD.\\
$^4$Mullard Space Science Laboratory, University College London,
Holmbury St Mary, Dorking RH5 6NT.\\
$^5$Email: lrj@star.sr.bham.ac.uk\\
}

\begin{document}
\maketitle

\begin{abstract}
We describe some of the first X-ray detections of groups of galaxies
at high redshifts (z$\sim$0.4), based on the UK deep X-ray survey
of \mch \etal (1998).  
Combined with other deep $ROSAT$ X-ray surveys with nearly complete optical
identifications, we investigate the 
X-ray evolution of these systems.  
We find no evidence for evolution of the X-ray luminosity
function up to z=0.5 at the low luminosities of groups of
galaxies and poor clusters (L$_X\gtrsim$10$^{42.5}$ \ergps), although
the small sample size precludes very accurate measurements.
This result confirms and extends to lower luminosities 
current results based on surveys at brighter X-ray fluxes. 
The evolution of the  X-ray luminosity function of these low luminosity 
systems is more sensitive to the
thermal history of the intra-group medium (IGM) than to cosmological 
parameters.  
Energy injection into the IGM (from for example 
supernovae or AGN winds) is required to explain
the X-ray properties of nearby groups. 
The observed lack of evolution suggests that the energy injection 
occured at redshifts z$>$0.5.  
\end{abstract}


\section{INTRODUCTION}

Groups of galaxies are ubiquitous; at least half of all local 
galaxies are located within groups (Tully 1987). 
A significant fraction of the total mass of the Universe may also 
be located in groups. 
In hierarchical models of the growth of structure in
the Universe, groups form before clusters and are the building-blocks
from which clusters assemble.

The evolution with redshift of galaxy groups has, however, received
little observational attention. This is largely because of the lack of
reliable samples of high redshift galaxy groups. Groups at high
redshift are difficult to find in two-dimensional optical surveys
because of their very low contrast against the foreground and background
galaxy distribution. Although surveys around radio galaxies have had
some success (eg. Allington-Smith \etal 1993), 
X-ray surveys have the potential to unambiguously
identify clusters and groups at high redshifts via the thermal X-ray
emission from the hot gas trapped in the cluster gravitational potential. X-ray
observations have shown that groups contain a hot intragroup medium
(IGM) which is a scaled down version of that found in clusters of
galaxies (Mulchaey \etal 1996, Ponman \etal 1996). This X-ray emitting
gas in groups may well constitute the largest observed component of the baryon
mass of the Universe  
(Fukugita \etal 1998).

The smaller  potential wells of groups, compared to clusters, imply
that the global X-ray properties of the inter-group medium (IGM) are
not only determined by the dark matter potential but are also  
sensitive to other heating (or cooling, Bryan 2000) processes
(eg. Cavaliere \etal 1997). Evidence has recently accumulated for 
non-gravitational 
energy input into the IGM. This energy is responsible for breaking 
the self-similar scaling
of cluster properties, most strongly in low mass systems 
(Kaiser 1991, Evrard \& Henry 1991).
Such additional heating can explain the shape of scaling relations
such as the steepening slope of the 
X-ray luminosity-temperature relation from clusters to groups
(Cavaliere \etal 1999,
Wu, Fabian \& Nulsen 1998, 2000, Balogh, Babul \& Patton 1999,
Bower \etal 2000, Tozzi \& Norman 2001, Lowenstein 2000).
The discovery of a minimum entropy, or `entropy floor', in the IGM
in the cores of low redshift groups has provided direct evidence of 
energy input in excess of that predicted from the scaled properties of
clusters (Ponman \etal 1999, Lloyd-Davies \etal 2000).
Possible origins of the 
additional energy include AGN or supernova-driven galaxy winds
at early epochs.
Studies of the evolution of the X-ray properties of groups can
discriminate between different origins (Valageas \& Silk 1999, Menci \& Cavaliere 2000).
 
Current X-ray surveys are sensitive enough to measure the ensemble 
properties of the local X-ray group population, but  
until recently X-ray surveys had insufficient sensitivity to detect
the low luminosities of galaxy groups at even moderately high redshifts
(z$\approx$0.3). 
Currently only 
the deepest $ROSAT$ surveys reach the fluxes
required ($\lesssim$10$^{-14}$ \ecs, 0.5-2 keV) and have sufficient optical
followup information. None of the current
$ROSAT$ cluster surveys (Jones \etal 1998, Rosati \etal 1998a, 
Vikhlinin \etal 1998, Burke \etal 1997, Romer \etal 2000) 
sample these faint fluxes. Here we report
on results based on the UK Deep $ROSAT$ Survey (\mch \etal 1998, hereafter M98). We
include updated positional information from a deep ROSAT 
high resolution imager (HRI) exposure.
We describe the groups and clusters found in the survey 
and combine them with other deep $ROSAT$ survey detections to 
measure their X-ray evolution. 
In Section 4 we show the Log(N)-Log(S) relation and X-ray luminosity 
function (XLF). The implications are discussed in
Section 5. We use H$_0$=50 km s$^{-1}$ Mpc$^{-1}$ and q$_0$=0.5 
unless otherwise stated.

\section{Observations}

\subsection{X-ray Observations~}

A detailed description of the observations made with the 
$ROSAT$  position sensitive proportional
counter
(PSPC) and of the 
optical follow-up observations are given in M98, so here
we provide only a summary, but also include a more detailed description of 
the $ROSAT$ HRI data.

The survey field is in a region of very low Galactic absorption
($N_H\approx6.5$x$10^{19}$ cm$^{-2}$) at position RA 13$^h$ 34$^m$
37$\fs$0, 
Dec 37\degr 54\arcmin
44\arcsec (J2000). The total PSPC exposure time was 115 ksec, making this survey the
second deepest $ROSAT$ X-ray survey made. Only the central region 
of the PSPC field, at off-axis angles $<$15 arcmin, was used. A total of
96 sources above a flux limit of 2x10$^{-15}$ \ecs (0.5-2 keV) were detected. 
The HRI observations were made in June 1997 and the total exposure
time was 201 ksec. The HRI field of view (38 arcmin square) covers the
complete survey area, and the spatial resolution is significantly
superior to that of the PSPC, ranging from 6 arcsec (half-energy diameter) on
axis to 30 arcsec (half-energy diameter) at 15 arcmin off-axis, 
although the narrow core of the HRI PSF  is maintained at these off-axis
angles.
The HRI detects X-rays over the full ROSAT energy
band (0.1-2.4 keV), but has lower sensitivity 
than the  PSPC,
because of its higher unrejected particle
background and lower quantum efficiency.
The HRI sensitivity is  particularly low 
to faint diffuse X-ray emission, for which  the improved 
HRI spatial resolution affords little gain in signal-to-noise.
In this work we are primarily interested in the extended X-ray emission
from groups and clusters of galaxies, and so we base our survey on the
more sensitive PSPC data, and use the HRI data to reduce error circle
sizes and help resolve confused sources.
We maintain the same source numbering system as used in M98.

\subsection{Optical, Radio and NIR Observations~}

Optical CCD imaging has been obtained  
 at the 3.6-m Canada-France-Hawaii
Telescope (CFHT),
the University of
Hawaii 2.2-m telescope, the 2.5-m Nordic Optical Telescope
and the Michigan-Dartmouth-MIT  2.4-m telescope. 
The deepest images, from the CFHT in sub-arcsecond seeing,  
reach R=24.5 mag over the whole field.
We have also made deep VLA radio 
maps at 20cm and
6cm, reaching a flux limit of 0.3 mJy at 20 cm. 
The $ROSAT$ positions were 
corrected for a  PSPC systematic position error (13\arcsec)
and the small ROSAT roll angle error of 0.185\degr (see M98).
Most of the X-ray sources were identified with objects brighter than R=23 mag.
Selected sources have near infra-red imaging (Newsam \etal 1997). 

Low resolution spectroscopy (10-15\AA) was performed at the 3.6m CFHT with the
MOS multislit spectrograph and at the 4.2m William Herschel Telescope (see 
M98 for details). Spectra were obtained within a contiguous region 
covering 85\% of the central 30 arcmin diameter survey area, and
containing 78 X-ray sources above the  flux limit of 
2x10$^{-15}$ erg cm$^{-2}$ s$^{-1}$ (0.5-2 keV). The survey region was largely defined
by the positioning of the MOS fields and did not include some areas at large off-axis angles.
Here we include the small region labelled `mask K' and not used by M98, in order to utilise
as large an area as possible. A total of 63 of the 78 sources have been
spectroscopically identified. The geometric area covered is 0.167 deg$^2$.

\section{Analysis and Results}

\subsection{Source Searching}

We have supplemented the PSPC source list of \mch \etal (1998), which was
produced using an algorithm optimised for point sources, with (a) HRI
information, and (b) a separate search of the PSPC data using the 
Voronoi tessellation and percolation (VTP) method (Ebeling \& Wiedenmann 1993),
which is optimized to locate faint extended sources.  

The HRI image was searched for point sources independently of the PSPC 
data using a maximum-likelihood fit to the point spread function (PSF),
including the radial variation of the PSF. The point source flux 
limit of the HRI observation was 
$\approx$3.5x10$^{-15}$
\ecs (0.5-2 keV), depending on the spectrum. 
A total of 43  HRI sources (of 61 detected) 
above a significance of 4$\sigma$ matched PSPC
sources. 
The fraction of PSPC sources detected in the HRI 
(at the 4$\sigma$ threshold)  falls 
as a function of flux because of the lower sensitivity of the HRI exposure
(see Table 1).
A full description of the impact of the HRI data on the total
identification
content of the survey is beyond the scope of this paper; here we
merely note that the HRI positions confirm that the PSPC
positions were accurate to within, or better than, the
statistical uncertainty given by M98, at least for PSPC fluxes 
$\gtrsim$3.5x10$^{-15}$ \ecs, where we have HRI detections.
In particular, the groups and clusters considered here 
all fall in the flux range ($>$5x10$^{-15}$ \ecs) where the HRI
positions of point sources have in general confirmed the PSPC positions.  

\begin{table}
\caption {HRI and PSPC coincidences}
 \begin{tabular}{lll} \hline
PSPC flux range & Number of HRI  & Median PSPC-HRI \\
(erg cm$^{-2}$ s$^{-1}$, & detections of PSPC  & position offset \\
0.5-2 keV) &     sources               & (arcsec) \\
  & &   \\
$>$1x10$^{-14}$ & 22 of 22 & 3.7 \\
5x10$^{-15}$-1x10$^{-14}$ & 12 of 18 & 5.7 \\
2.9x10$^{-15}$- 5x10$^{-15}$ & 8 of 37 & 4.0 \\
  & &  \\

             \hline
\end{tabular}
\normalsize
\end{table}

The Voronoi tessellation and percolation algorithm was applied to the 
PSPC data,
as described by Scharf \etal (1997), to check if the point-source optimized 
algorithm of M98 had missed any
extended sources. Only one such source was found (number 173) and it 
was added to the list of candidate clusters.

PSPC count rates in the 0.5-2 keV band were measured for all group/cluster 
candidates within an aperture of radius 40-60 arcsec (270-410 kpc at
z=0.4), after masking out any
nearby
sources and using a local
background subtraction and an exposure map created using the Snowden
\etal (1992) method. These count rates were in general higher than those
measured by M98 assuming point-like sources, but agreed to within 
$<$10\% with the available VTP count rates. To convert from count rate to 
(absorbed) flux we used a constant value of 1.15x10$^{-11}$ \ecs
(0.5-2 keV) (ct/s)$^{-1}$. This conversion is accurate to within 5\% for 
temperatures in the range $\approx$1-2 keV (the temperatures expected from the 
luminosity-temperature relation for the luminosities we find
eg. Fairley \etal 2000) given the low 
Galactic column density and assuming abundances of 0.3 times solar. 
The correction for Galactic absorption in the direction of this 
field is very small ($\la$2\% in the 0.5-2 keV band for a 
temperature of 1 keV, and less for higher temperatures) and has been
assumed to be negligible.  

A correction for missing flux outside the measurement aperture can
only be made approximately because the PSPC 
surface brightness profiles are not measured accurately enough to
allow extrapolations to larger radii.
Assuming  a King profile with $\beta$=2/3 and a core radius of
100 kpc, appropriate for groups (although there is a large scatter in
observed values eg. Mulchaey \etal 1996) the missing flux, calculated
separately for each cluster,  was typically 
30\% at z=0.3
and 23\% at z=0.5. The final fluxes and K-corrected luminosities
are given in Tables 1 and 2. 
The K-corrections and bolometric luminosities were derived in a 
self-consistent manner using temperatures estimated from an L$_X$-T relation.
The K-corrections for the estimated  
temperatures  of 1-2 keV  
were in general small ($\la$10\%; see
Jones \etal 1998). However, given the relatively low signal-to-noise 
ratios of the X-ray detections (in the range 6-10), the scatter
in possible values of $\beta$ and the core radius, and the unknown
temperatures, we estimate that the fluxes and luminosities
are accurate to no better than $\approx$40\%.

\subsection{The group and poor cluster sample}

The properties of all the group and cluster
identifications within the complete survey area are given in 
Table 2. Other sources, which are 
candidate groups or clusters, but for which we have insufficient information
to be sure, are listed in an appendix.
With new HRI data we confirm the hypothesis noted in M98 that
source 5 contains a contribution from an AGN as well as from a possible
cluster. A point-like, variable, source dominates the HRI emission. 
The remaining faint extended emission (source 5b)
may be associated with a distant cluster of unknown redshift (see below).

Significantly, none of the 5 group or cluster identifications were detected
in the HRI image. A visual inspection of the smoothed HRI image
confirmed the non-detections. 
In fact, of all the 96 PSPC sources, the brightest {\it not} detected in the HRI 
(source 34) is identified with a group/cluster. Six of the 18 
PSPC sources with PSPC fluxes 
between 5x10$^{-15}$ \ecs and 1x10$^{-14}$ \ecs
are undetected in the HRI. Of these 6,
3 (or possibly 4) 
are optically identified as groups/clusters, and the
others are probably X-ray variable AGN.
Because of the poor HRI sensitivity to faint diffuse emission,
this result is expected, and supports the identifications with groups/clusters.

In order to confirm the likelihood of non-detection in the HRI, a simulation
was performed of an extended source with the PSPC total flux of R34,
the brightest source in the sample. 
A King profile with core
radius of 0.5 arcmin (or $\approx$200 kpc at the redshift of R34),
$\beta=2/3$, and a temperature of 2 keV were assumed. Fluctuations
based on Poisson 
statistics were included in the
simulation. The total number of HRI counts was $\approx$150, but since the
background level was 630 ct arcmin$^{-2}$ (the same 
as in the real HRI image), the maximum significance of the source was 3.1
sigma, within a radius of 30 arcsec, and less significant at larger radii.
Thus such a source would fall just below the threshold of significant detection.

\begin{table*}
\begin{minipage}{175mm}
\caption {Properties of the groups and clusters}
\tiny
 \begin{tabular}{lllllllllllllll} \hline
ID & \multicolumn{1}{c}{PSPC (0.5-2 keV)}& \multicolumn{2}{l}{PSPC
position (J2000)} &  
Offset & Id & Hardness & HRI & R & n$_z$  &  z & \multicolumn{2}{c}{$L_{X,42}$} 
& $N_{0.5}^c$ & PSPC \\
 & flux x10$^{-15}$& RA &Dec & 
(\arcsec) & class & ratio& detect & mag & &   & 0.5-2 & bol & & extended?\\
  & & & & & & & & & & & & &  \\
  & [a] & [b]&[c]&[d]&[e]&[f]&[g]&[h]&[i]&[j]&[k]&[l]&[m]&[n] \\
  & & & & & & & & & & & & &  \\

34 &21.4   & 13 35 14.07 & 37 49 01.7 & 6.4 & * & 0.86$\pm$0.15& No &
21.3 &2 & 0.595$\ddag$ &38 & 93& 5.4$\pm$3.1 &Yes\\
58 &17.2   & 13 34 34.00 & 37 57 03.2 & 7.7 & * & $>$0.90      & No &
19.0 &2 & 0.308 &8.2 & 17 & 10.3$\pm$3.0&Yes \\

74 &7.2   & 13 34 07.63 & 38 06 20.2 & 6.6 & * & $>$0.66      & No &
18.8 & 2 & 0.382 &5.6 & 11& 3.1$\pm$1.5 & \\
96 &7.7   & 13 34 58.47 & 37 50 23.2 & 11.8 & (*)& $>$0.53 & No &
19.2 & 1 & 0.382 &6.0 & 12 & 21$\pm$5.5 & \\
173$\dag$&8.7   & 13 33 30.0 & 37 54 55.0 & 10 & (*)    &  -  &  No &
20.2 & 5 & 0.383 &6.7 & 14 & 11.2$\pm$4.5  &Yes \\

             \hline
\end{tabular}
\normalsize

(a) PSPC total flux in units of 10$^{-15}$ \ecs (0.5-2 keV), corrected for 
flux outside the measurement aperture.\\ 
(b) \& (c) X-ray position from PSPC data.\\
(d) Offset of the X-ray position from the optical counterpart.\\
(e) Reliability of the identification, based on all available data,
from M98 but updated using HRI data for 3 sources: 
`*' means certain, `(*)' means likely and
blank means possible.\\
(f) Ratio of the counts in the 0.5-2 keV band to those in the 0.1-0.5 keV band
(see M98 for details). \\
(g) HRI detections. `No' means undetected and `pt' means a point-like
source.\\
(h) R band magnitude of the brightest cluster galaxy or other counterpart\\
(i) Number of measured and concordant galaxy redshifts.\\
(j) Redshift\\
(k,l) X-ray luminosities in units of 10$^{42}$ erg s$^{-1}$, for H$_0$=50 
km s$^{-1}$ Mpc$^{-1}$ and q$_0$=0.5: (k) 0.5-2 keV, (l) bolometric,
based on thermal spectral models (see text).\\
(m) Richness as defined by Bahcall (1981).\\
(n) Whether a source was detected as extended in the PSPC.\\
$\dag$ Source 173 was not in the original M98 0.5-2 keV sample 
(although it is mentioned
there in the notes to source 43) and was found using VTP.\\
$\ddag$ Redshift uncertain.

\end{minipage}
\end{table*}

Recently obtained $Chandra$ and $XMM-Newton$ data
on the UK deep field confirm that sources 34, 58 and 96 are indeed extended 
X-ray sources with little point-source contamination.
Source 74 may however be contaminated by a point-like X-ray source;
source 173 is at the very edge of the field and is a marginal detection. 
A detailed investigation of these data will be presented in a future paper.

The X-ray luminosities of the sources in Table 2 range from 5x10$^{42}$
\ergps
to 4x10$^{43}$ \ergps (0.5-2 keV). This luminosity range is that of groups of galaxies 
up to poor clusters (eg. Mulchaey \etal 1996, Ponman \etal 1996). 

The optical richness of each system in Table 2 was measured from our R band
photometry. The $N_{0.5}^c$ parameter of Bahcall (1981) was used as a 
measure of the galaxy overdensity within 
a radius of 0.5h$_{50}^{-1}$ Mpc. A local galaxy background was subtracted for 
each system, and a richness correction applied as described by Bahcall (1981).
The errors quoted in Table 2 are based on 
Poisson statistics only. $N_{0.5}^c$ is not an accurate measure 
of richness for poor systems such as these, given the large number of 
projected foreground and background galaxies, and only 2 of the 5 systems 
have $N_{0.5}^c$ detected above 3$\sigma$ significance. The low values
of $N_{0.5}^c$=3-20 reflect
the group and poor cluster nature of the systems; Abell richness class
0-1 clusters typically have  $N_{0.5}^c$=10-50.

\subsection{Sample Completeness} 

The sample of five reliably identified groups and low luminosity clusters
all have fluxes $>$7x10$^{-15}$ \ecs (0.5-2 keV) and are thus in
the flux regime where HRI data have confirmed the PSPC positions of 
point X-ray sources. Three of the five are extended sources
in the PSPC data and four have high hardness ratios consistent with
the predictions for thermal spectra given in Table 6 of M98. 
The number of confirming galaxy redshifts per group 
ranges from 1 to 5 (see Table 1).

We have identified a further 8 candidate groups and clusters (see the
appendix), although
one (5b) is of low flux, two contain HRI point-like X-ray sources and thus may
be dominated by AGN emission, two are predicted to be at z$>$1.3,
and two are outside the complete survey area.  
In addition, 
there are several unidentified sources in the catalogue of M98
which could be clusters. However, these sources are mostly unidentified
because of their faint optical or near infra-red counterparts 
(Newsam \etal 1997) and thus are likely to be at much higher redshifts
(and higher luminosities) than considered here 
(eg. one of the unidentified sources is R112, listed in the appendix 
as a z$>$1.3
cluster candidate). 

We conclude that the  sample
of five groups and clusters is a lower limit to the true 
number, but that for redshifts z$<$0.7 it is reasonably free of 
uncertainties due to incompleteness. 

We use a limit in total flux of 6x10$^{-15}$ \ecs (0.5-2 keV), higher
than that used by M98 for point sources, 
in order to be reasonably sure of not missing
extended low surface brightness systems.  
At this flux limit, sources could be detected over the full 
geometric sky area of the survey unless they exceeded a 
critical angular size, and thus had a central surface brightness below
the background-limited threshold. Assuming a King profile, and
within a radius of 45 arcsec, corresponding to our measurement apertures
(300 kpc or $\approx$2 core radii at z=0.4), the  surface 
brightness within 2 core radii of sources at the flux limit is detectable
at all redshifts z$>$0.1. In the 0.3$<$z$<$0.7 range used for the
X-ray luminosity function in Section 5, the sample should be flux limited,
rather than  surface brightness limited. At z$<$0.1, the large angular size (r$>$1.5 
arcmin) of sources  at the flux limit make them undetectable in our data;
however these sources would have luminosities L$_X\lesssim$3x10$^{41}$ \ergps,
well below the luminosities studied here.

\subsection{Notes on individual sources}

Images of all the confirmed groups and clusters are shown in Fig. 1.

\noindent{\bf 34.}
The PSPC X-ray data are significantly extended over $\approx$1.5 arcmin
and the X-ray hardness ratio is high, consistent with a thermal spectrum (see
Table 6 of M98).   
A compact group of at least 6 faint galaxies lies within the 10 arcsec radius PSPC
error circle (see Fig. 1). The optical spectra are of insufficient quality to
be completely certain of the redshift, but  two of the galaxy spectra
show features consistent with Ca H \& K and G band absorption lines
at z=0.595. The faint magnitudes of the central galaxies are consistent
with this relatively high redshift. The X-ray luminosity (4x10$^{43}$
\ergps) implied by this redshift is that of a poor cluster of galaxies, but 
the optical appearance
of the cluster core is more like that of a compact group, such as
Hicksons compact groups. The X-ray luminosity of this source is, however,
 higher than the luminosity of the diffuse emission in any 
of the Hickson compact groups studied by Ponman \etal (1996),
although here we include in the luminosity any emission from the
cluster galaxies. 
The galaxies in the core have the absolute magnitudes
of L$^*$ galaxies, and some of them may be in the process
of merging to form a luminous central galaxy. No central giant elliptical
galaxy is currently observed.

\noindent{\bf 58.}
The PSPC data show an extended, hard X-ray source (or possibly a complex
region of at least 3 point sources). Two (early-type) 
galaxies near the X-ray centroid
have identical redshifts of z=0.307$\pm$0.001. A third (spiral) galaxy $\approx$1 arcmin
further north also has a similar redshift of z=0.31.

\noindent{\bf 74.}
Although no X-ray extension is detected, the hardness ratio is high and
the X-ray centroid is coincident 
with an excess of galaxies within a
region of radius $\approx$30 arcsec. Two of the galaxies nearest the X-ray
centroid have identical absorption line redshifts of 0.382. 
Recent $Chandra$ data indicate that a significant fraction of the quoted X-ray 
luminosity may arise in a point X-ray source, and further investigation
is necessary. However, there is still a probable optical group at this location.
The group redshift is very
similar to the redshifts of sources 96 and 173, both identified with 
a group or cluster. The three sources are 13-16 arcmin apart, or 4.9-6.0 Mpc
at z=0.382.

\noindent{\bf 96.}
A poor cluster is apparently coincident with the X-ray position, as confirmed by
the value of $N_{0.5}^c$=21$\pm$5.5. The X-rays show no extension,
possibly due to the low X-ray flux.
The redshift of the
brightest cluster galaxy is 0.382. This galaxy is a giant elliptical with 
$M_R$=-23.2 which hosts an extended radio source. However, no point X-ray source, 
indicative of AGN activity, was detected in the HRI exposure.

\noindent{\bf 173.} This extended X-ray source was detected by the VTP algorithm 
and not listed by M98 because it was not reliably detected in the 0.5-2 keV band 
by the point-source optimized algorithm used by M98.
Sources 173 and 43 are only one arcmin apart and are merged by VTP. 
We have 5 concordant galaxy redshifts in this area
including the counterpart to source 43, as listed by M98. 
Although source 43 was
not detected at a significant level in the HRI, a faint X-ray source (at 
$\approx$2.6$\sigma$ significance)
is present at the position of the  
counterpart galaxy, which  has narrow emission lines and is a bright double radio source
(the brightest in the ROSAT field). Together with the X-ray contours and optical
appearance shown in Fig 1, this suggests that source 43 is an outlying AGN in the group,
and that the emission from 
extended source 173, centred near to the brightest galaxy of the group,  
originates in the hot X-ray gas.    
The flux was measured in an aperture of radius 36 arcsec centred on source 173, 
which excluded the flux from source 43.

\begin{figure}
\psfig{figure=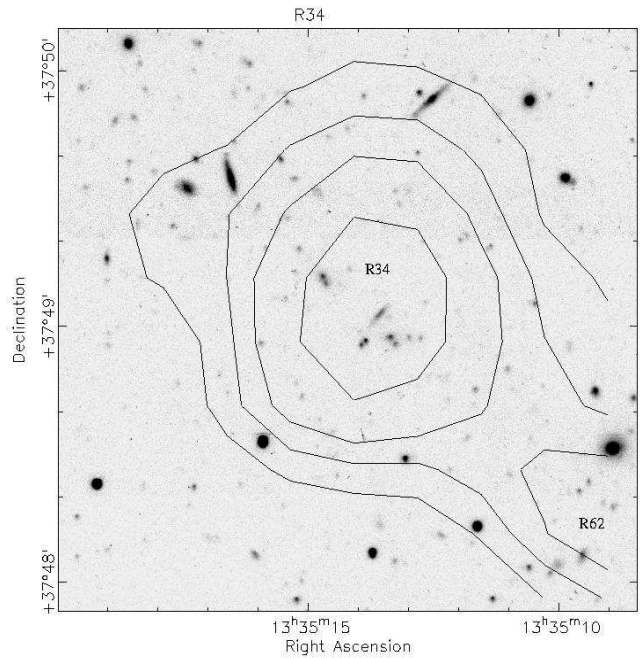,height=13.0truecm,angle=0}
\caption{$R$-band CCD images and overlaid X-ray contours from the $ROSAT$ PSPC (0.5-2 keV)
and from the $ROSAT$ HRI (dotted lines) of the groups and clusters in the UK deep
field.
The X-ray contours are at logarithmic intervals of a 
factor of 1.2  in surface brightness.
}
\end{figure}
\begin{figure}
\psfig{figure=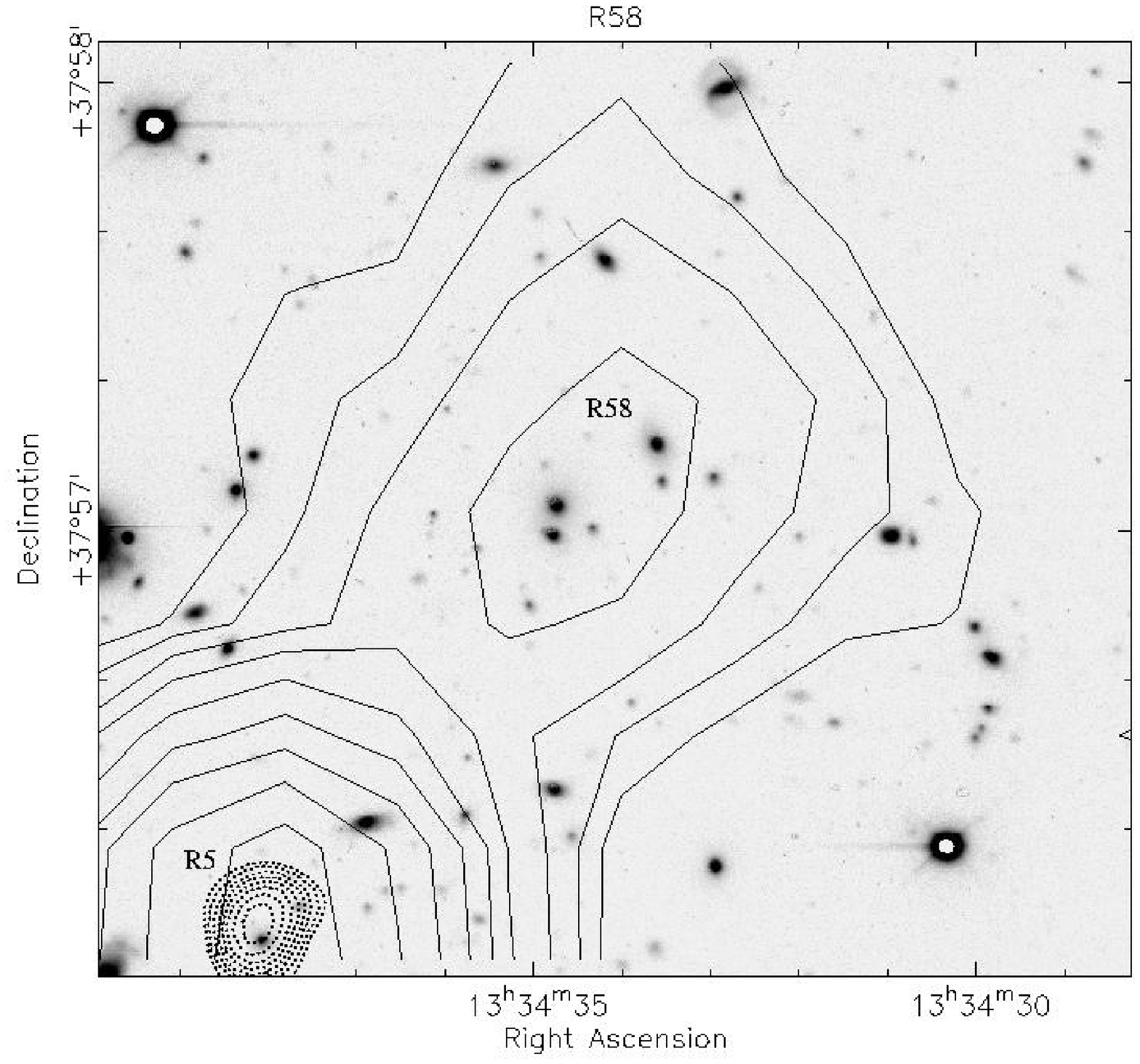,height=13.0truecm,angle=0}
\end{figure}
\begin{figure}
\psfig{figure=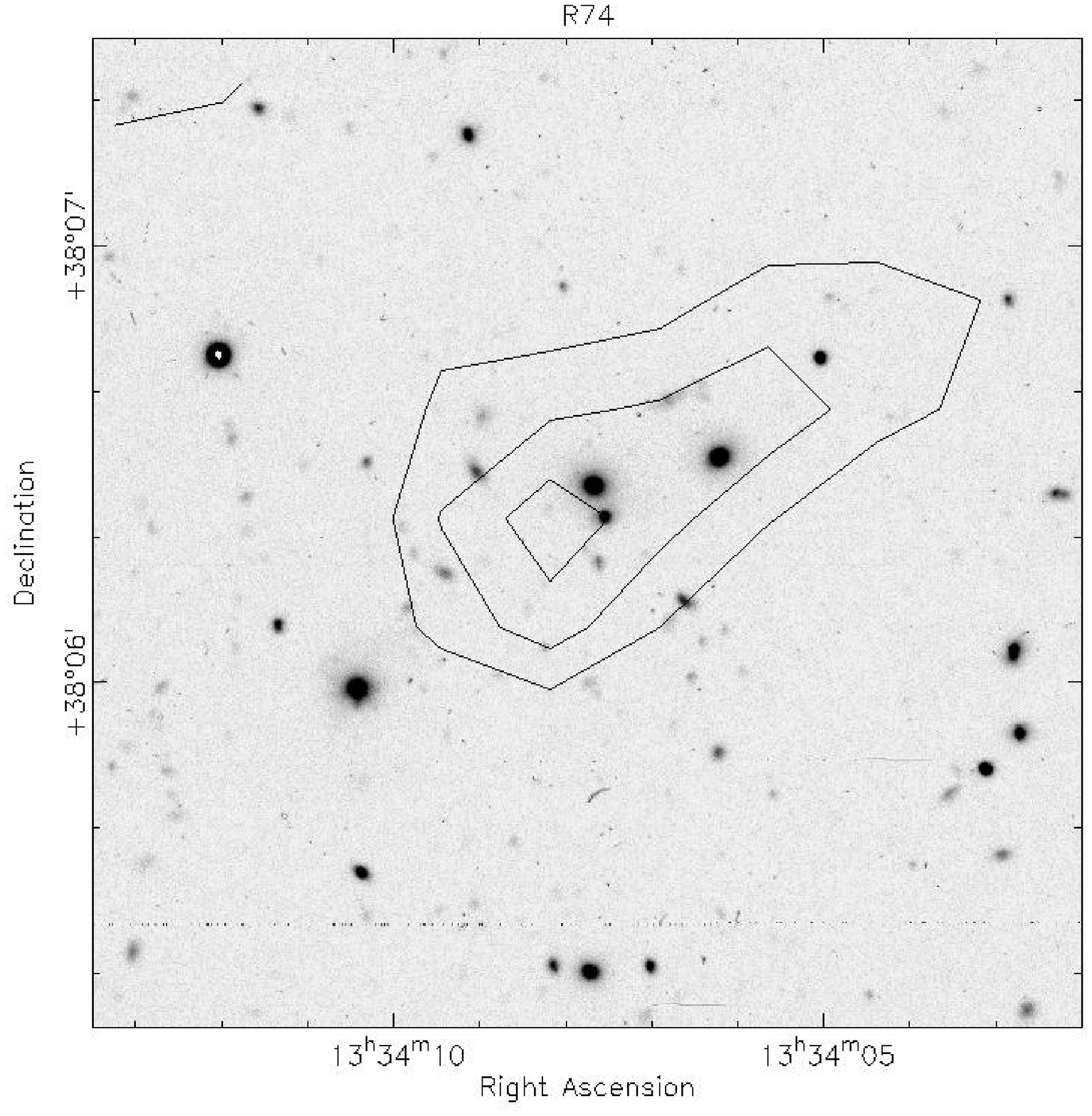,height=13.0truecm,angle=0}
\end{figure}
\begin{figure}
\psfig{figure=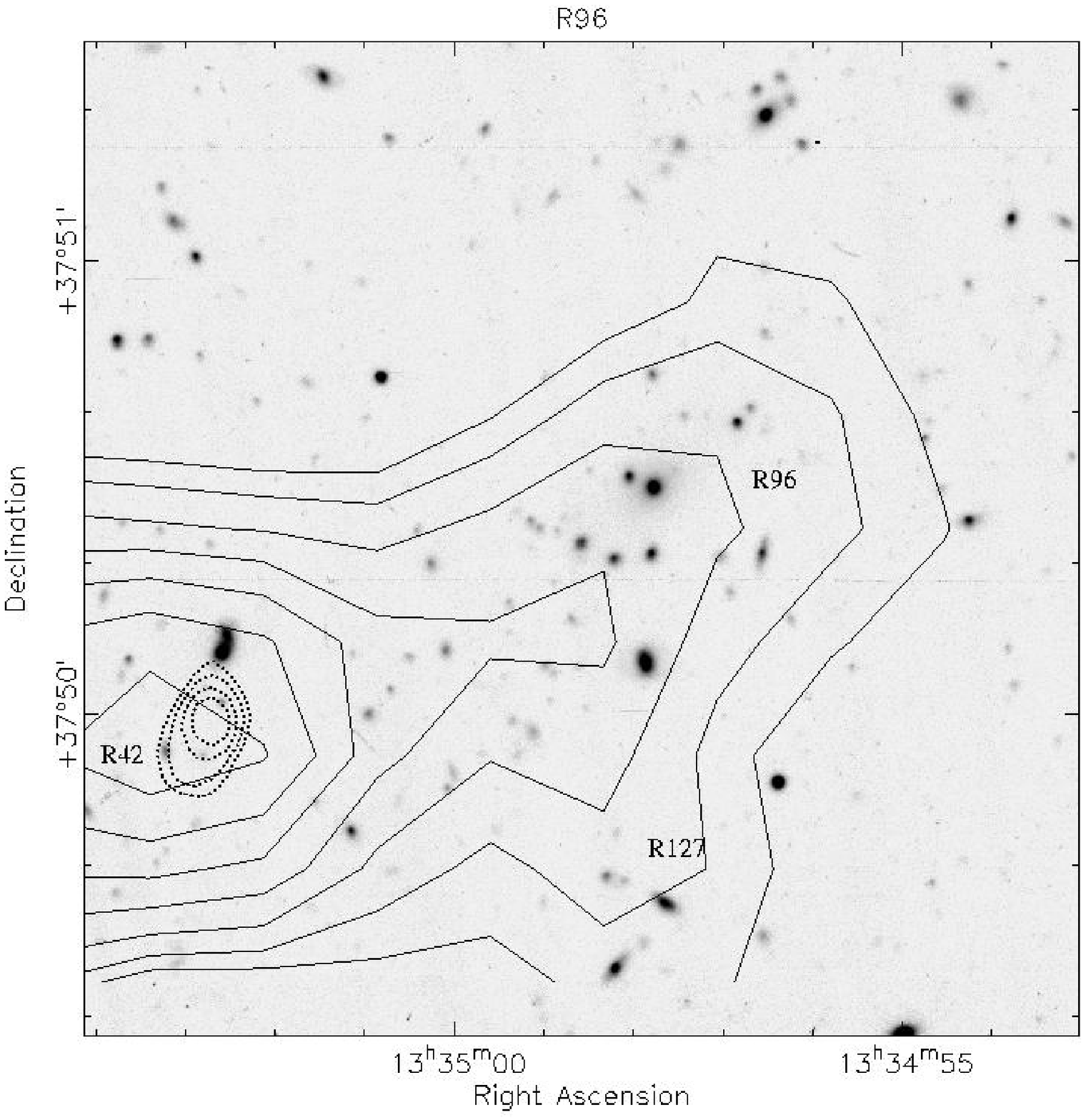,height=13.0truecm,angle=0}
\end{figure}
\begin{figure}
\psfig{figure=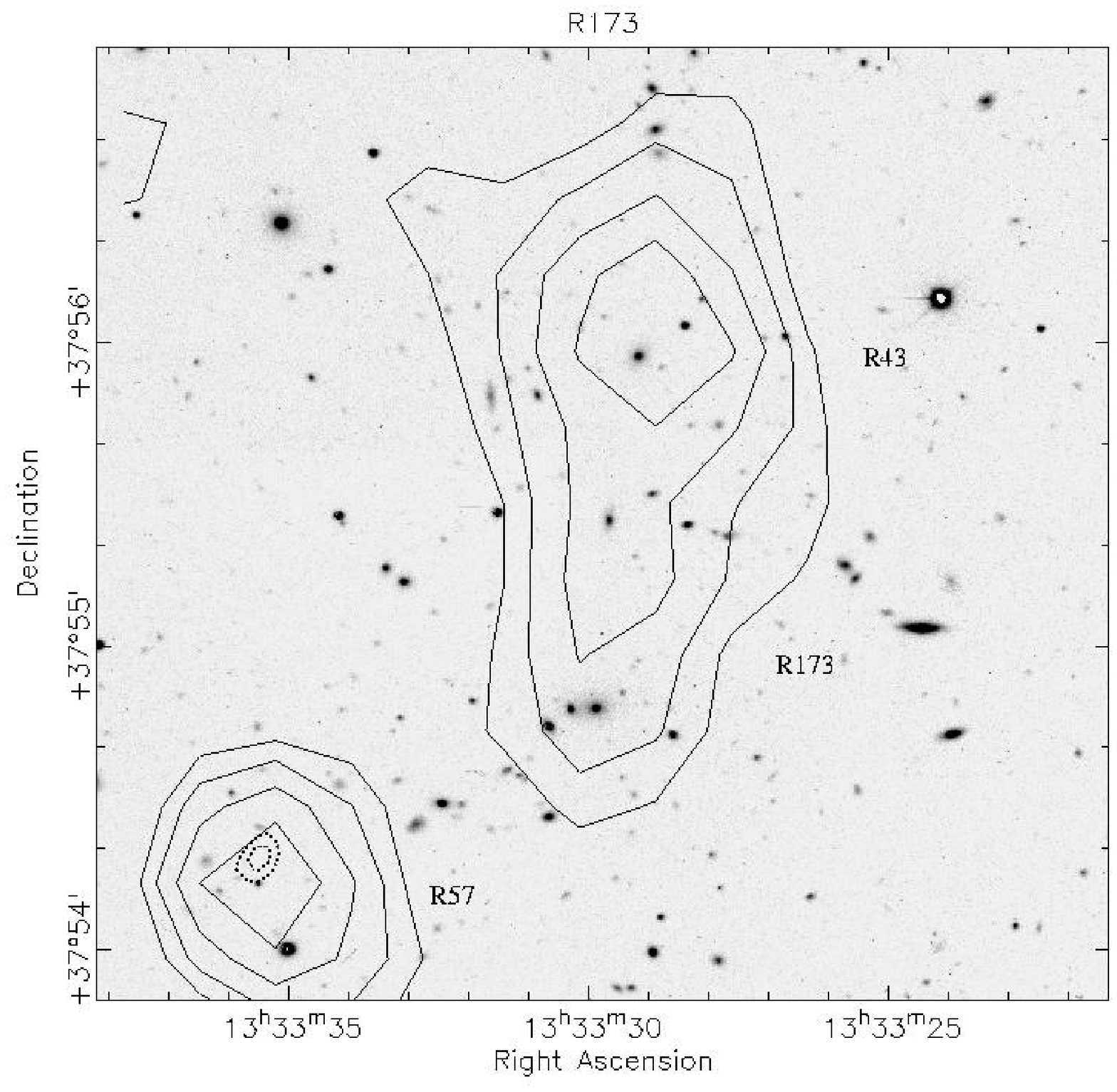,height=13.0truecm,angle=0}
\end{figure}

\subsection{Clusters of unusually low X-ray luminosity}

During an inspection of the R band optical images obtained for the
survey, several candidate distant clusters of galaxies unrelated to any
X-ray source were noted. These clusters 
were optically selected, by eye, and in no way form a complete sample. 
For one cluster, we have a redshift of z=0.519 based on two
galaxies with concordant redshifts 
(cluster G; RA 13$^h$ 34$^m$ 05$\fs$6, Dec 37\degr 51\arcmin 41\arcsec,
J2000),
allowing us to place an upper
limit on the total X-ray luminosity of the cluster of L$_X<$8x10$^{42}$
\ergps (0.5-2 keV). The cluster is only 45 arcsec from QSO number 101 of M98;
allowing for the possibility of confusion has resulted in a
relatively high upper limit. 
Without further redshifts, we cannot be certain that this system is a
real cluster and not just a projection effect.
Its detectability by eye is a result of it being dominated by
a luminous elliptical galaxy.
Its apparent optical richness is not greatly higher than that of
the other X-ray selected groups and poor clusters and its X-ray
flux may be just below the survey limit. However, the upper limit
on the X-ray luminosity is lower than found in other studies 
of optically selected clusters (eg. Bower et al 1994). 
It may be a cluster in the early stages of formation. However, if it is a real, 
bound, system and its 
X-ray luminosity is unusually low, then
some other interesting
possibilities are raised, including
an unusually high efficiency of galaxy formation, leaving little
primordial gas behind, a large injection of energy from AGN or 
supernovae-driven winds (see Section 5), or a low ICM metal abundance, 
reducing the soft
X-ray luminosity. Further observations are required to clarify
the situation.

\subsection{Data from other $ROSAT$ surveys}

Results based on observations of only
one narrow field are subject to biases due to large scale-structure
in the direction of that field. For this reason we have compiled additional 
group/cluster detections from the deep $ROSAT$ PSPC surveys 
of Hasinger \etal (1998a), Zamorani \etal (1999) and Bower \etal (1996). 
For consistency, we
only use deep surveys based on PSPC data. In all these
surveys a  considerable effort has been made by the authors to identify all of
the X-ray sources.  

The Lockman hole survey of  Hasinger \etal (1998a), Schmidt \etal (1998),
\& Lehmann \etal (2000) was obtained using 
207 ksec of PSPC data combined with HRI data.  
A source detection algorithm employing a maximum likelihood fit to
the PSF was used. All 50 X-ray sources with fluxes $>$5.5x10$^{-15}$ \ecs (0.5-2 keV) 
have been identified, resulting in 3 group/cluster identifications.
All 3 (numbers 41, 58, 67) are extended X-ray sources with multiple 
galaxy redshifts per system.   
We do not include the high redshift clusters reported in 
Hasinger \etal (1998b) because these were outside the complete survey
area, or below the flux limit, of the  Hasinger \etal (1998a) PSPC survey.

The survey of Zamorani \etal (1999) used a PSPC exposure of
56 ksec on the Marano field. Fifty X-ray sources were detected 
above a flux limit of 3.7x10$^{-15}$ \ecs (0.5-2 keV) using the same
source detection algorithm as Hasinger \etal (1998a). Of the 42 sources
identified, 3 were clusters or groups of galaxies with up to 3 galaxy redshifts per
system. None of the unidentified sources are coincident
with an excess of galaxies on R band CCD images at R$\lesssim$24,
making their identification with clusters or groups at z$<$0.7 
very unlikely. One of the 3 clusters has no spectroscopic redshift;
we estimate a redshift of z$\approx$1.1 based on the magnitude of R=22.1 of the
brightest galaxy given by Zamorani \etal (1999).

In the deep (79 ksec) NEP survey of Bower \etal (1996) identifications were 
made for 18 out of 20 X-ray sources brighter than 1x10$^{-14}$ \ecs (0.5-2 keV),
including one cluster. 
The diffuse extended source in this field identified by Burg \etal (1992) as
a nearby cluster (NEP X1) at z=0.088 has also been included. We note that by minimizing
the size of the PSPC PSF  (by using the hard 0.5-2 keV band)
and including the extra exposure of Bower \etal (1996), NEP X1 breaks
up into several components, some of which are consistent with point sources
(this can be seen to some extent in Fig 1 of Bower \etal). 
Thus the true extent of the diffuse emission is probably considerably less than 
the 20x10 arcmin$^2$ claimed by Burg \etal The excess X-ray emission concentrated
in this part of the field may be partly due to AGN, perhaps associated with the
group. We have extrapolated from the flux within a radius of 2.0 arcmin 
(270 kpc at z=0.088) to derive
a total flux of 5x10$^{-14}$ \ecs and luminosity of $\approx$2x10$^{42}$ 
\ergps (0.5-2 keV) for NEP X1. 

We have remeasured the fluxes of the groups/clusters
in these additional fields from archival $ROSAT$ PSPC data using the
same method as for the UK Deep survey field, except that in several (higher signal-noise) 
cases we used a larger aperture, up to 85 arcsec radius. 
The limiting total flux adopted here was taken from the published limit for
identifications of point X-ray sources, multiplied by the mean ratio
of total flux to published flux for the group/clusters in each survey.
This ratio was 1.5-1.6 for all three surveys (see Table 3). The Lockman Hole
survey has two flux limits, corresponding to off-axis angles $<$12.5 arcmin
and 12.5 to 18.5 arcmin.
We have applied the VTP 
algorithm to the NEP field to check for undetected low surface brightness 
extended sources.
Two additional candidates were found, but one of these is NEP X1, already 
included, and the other, although marginally extended, 
is consistent with most of its flux originating in a
point-like X-ray source.
We thus expect that at most one or two extended sources at moderate redshifts 
will be missing.

\subsection{Limitations of the data}

Given that the number of groups/clusters in some
fields is smaller than
the number of unidentified sources, the group/cluster numbers we derive 
are strictly lower limits. However, the unidentified sources are usually
lacking optical counterparts bright enough for spectroscopic identification,
and if they are clusters, they are thus likely to be at z$\gtrsim$1
(eg. source R112 in the UK deep field and the tentative z$\sim$1.1 cluster
identification in the Marano field). They are almost certainly
at z$>$0.7, the upper limit used here.  

The X-ray data are of insufficient resolution and signal-noise to allow
a correction for the contribution of individual galaxies and AGN 
to the X-ray luminosities, 
so the true IGM luminosities are probably
lower than the values given here, particularly for the lowest luminosity
systems. In our statistical measurements, we compare with 
low redshift X-ray data which were also not corrected in this way, so
the comparison is a reasonably valid one. However, AGN evolution
of the form $L_X\propto$(1+z)$^3$ implies a higher rate of contamination
at higher redshifts, which we have not corrected for.

\begin{table}
\caption {Groups and clusters in deep $ROSAT$ PSPC surveys}

 \begin{tabular}{llll} \hline
Field & area & f$_{X_{tot,lim}}$ & no. groups \&  \\
      & (deg$^2$) & (0.5-2 keV) & clusters (z$<$0.7)  \\
  & & &    \\
(a)& &(b) & (c) \\
  & & &     \\

UK deep & 0.167 & 6.0x10$^{-15}$ & 5  \\
Marano (Z99)  & 0.20 & 6.4x10$^{-15}$ & 2  \\
NEP (B96)    & 0.21 & 16x10$^{-15}$ & 2  \\
Lockman (H98;S98)& 0.14 & 8.4x10$^{-15}$ & 3  \\
~Hole    & 0.16 & 17x10$^{-15}$ &   \\
             \hline
\end{tabular}
\small
(a) Z99=Zamorani \etal (1999); B96=Bower \etal (1996); H98=Hasinger \etal (1998);
S98=Schmidt \etal (1998)\\
(b) Limiting {\it total} flux adopted here for extended sources 
(ie. including flux 
outside the measurement aperture) in \ecs\\
(c) Number of confirmed group/clusters at z$<$0.7 \\
\normalsize
\end{table}

\section{Results}

\subsection{Log $N$-Log $S$ relation}

In Fig. 2 we show the binned integral \logn of the various deep 
surveys as well as $ROSAT$ cluster surveys at brighter fluxes.
The group/cluster number density of $\approx$20-30 deg$^{-2}$ is the highest 
measured in any current X-ray survey and is similar to that found in 
deep optical searches (eg. the number density of cluster candidates
found by Postman \etal 1996 was 16 deg$^{-2}$
at I$<$22.5 mag and estimated redshifts 0.2$<$z$<$1.2),
but X-ray surveys do not have the problems caused by projection effects.  
The large error bars reflect the small numbers of group/clusters in each
survey, but in general they are consistent with each other and
with the faintest point in the \logn of Rosati \etal (1998b),
who have located clusters via their extended X-ray emission in
a larger number of shallower $ROSAT$ PSPC fields than
used here. 
The deep survey points include a small additive correction of 3 deg$^{-2}$ 
for clusters missed at bright fluxes $>$4x10$^{-14}$ \ecs
because the area of sky sampled was too small to detect bright clusters.

The field with the highest surface density of groups and
clusters is the UK Deep field. The probable reason for this can be seen in
Table 2. Of the five systems in this field, three are at almost the 
same redshift and
are separated by $\approx$5 Mpc, suggesting that there is large-scale
structure in this direction at z=0.382. To account for this effect,
we have constructed the log$N$-log$S$ relation (Fig. 2) and 
the X-ray luminosity function (Fig. 3) in two ways, both excluding two of the 
z=0.382 systems in the UK Deep field (shown as solid circles) 
and including all three z=0.382 systems (solid squares). 
The two systems removed were those with the least reliable identifications
(96 and 173) but in practice which two are removed makes little
difference as all three have very similar fluxes and luminosities. 

Only one of the
other X-ray sources of M98 is close in redshift to these three
systems, so any   large-scale
structure does not appear to effect the redshift distribution of 
the AGN detected by $ROSAT$.

The no-evolution predictions in Fig. 2 are based on an integration of the 
local luminosity function of Ebeling \etal (1997) for luminosities  
$L_X>10^{42}$ \ergps and redshifts up to z=1.5. K-corrections based on 
an L-T relation are included.
The counts are consistent with no evolution. In the next section we 
include redshift information to test whether the luminosity functions
are also consistent with no evolution.

\begin{figure}
\psfig{figure=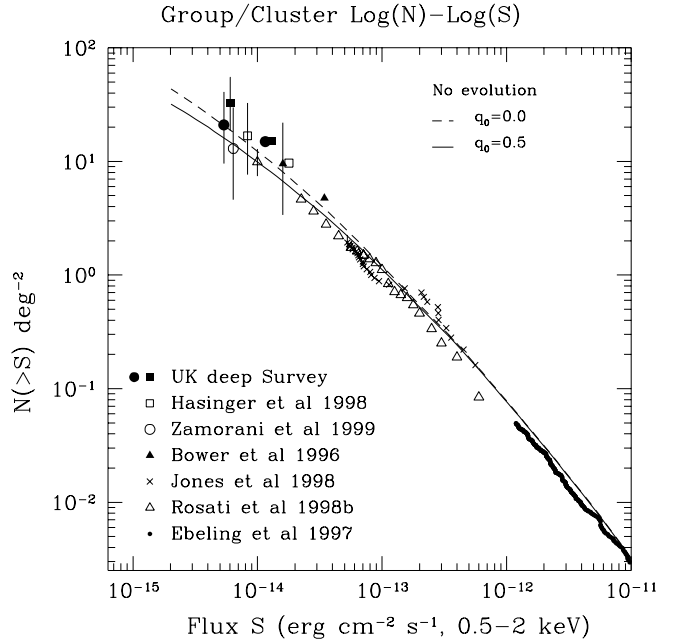,height=9.5truecm,angle=0}
\caption{
Integral \logn for X-ray selected groups and clusters in various deep 
$ROSAT$ surveys. No evolution predictions are also shown. Error bars
are only shown for the faintest point of each survey, since the data
points within each survey are not independent. For the UK deep survey,
the solid circles show the effect of removing an excess of two groups 
due to possible large-scale structure (and are offset along the x-axis 
for clarity).
}
\end{figure}

\begin{figure}
\psfig{figure=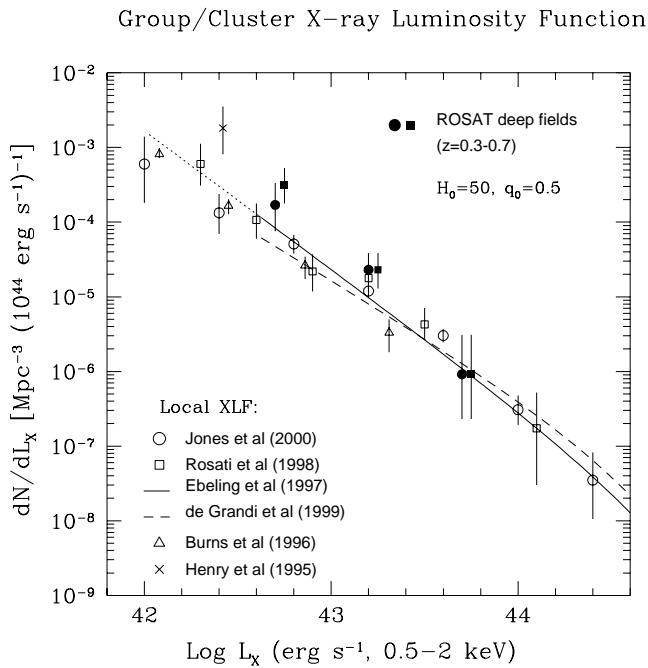,height=9.5truecm,angle=0}
\caption{
Evolution of the X-ray luminosity function at the low luminosities 
of groups \& poor clusters. The combined high redshift data from four 
$ROSAT$ deep surveys are shown
for all groups \& clusters  (filled squares) and without 
the possible excess of groups
due to large-scale structure in the UK Deep field (filled circles,
offset along the x-axis for clarity).
A variety of local measurements (z$<$0.3) are also shown.  
}
\end{figure}

\subsection{X-ray luminosity function}

The X-ray luminosity function of the eleven groups and clusters 
at 0.3$<$z$<$0.7 from
all four surveys combined is shown in Fig. 3 (filled squares).
The mean redshift is z=0.46 
and the redshift range is z=0.31-0.63. 
The filled circles in Fig. 3 show the effect of removing two of the three
groups at z=0.382 in the UK deep field, under the 
assumption that their presence is due to large-scale structure.
Comparison with local
luminosity functions, which were 
converted to the 0.5-2 keV passband and to H$_0$=50 where
necessary, shows that these data
are consistent with no evolution (or a small degree of positive
evolution) of the X-ray luminosity function   
of groups and poor clusters to redshifts z$\sim$0.5.

The discrepant Henry \etal (1995) point, whilst notable for being the first
statistical determination of  group X-ray properties, was based
on a sample size of only 3 groups and may have been affected by large-scale 
structure.  
The deepest previous 
survey with measured redshifts is that of Rosati \etal (1998a,b) who
did not sample luminosities below 10$^{43}$ \ergps at z$>$0.25.

The luminosity function was calculated using the 1/$V_a$ statistic 
of Avni \& Bachall (1980).
Luminosities were calculated in an iterative manner using the 
L-T relation of White \etal (1997) to estimate temperatures and hence 
derive K-corrections. Search volumes were derived using 
K-corrected luminosities and the limits in
total flux and survey areas given in Table 3.

\section{Discussion}

This is a small sample of groups and poor clusters but it probes 
lower luminosities than previous studies at high redshifts
(z$>$0.3).  We find no evidence for evolution of the X-ray luminosity 
function at the luminosities of groups and poor clusters 
($L_X\lesssim$10$^{43}$ erg s$^{-1}$), up to redshifts z$\approx$0.5,
confirming and extending the results of other X-ray cluster surveys.

The global evolution of the number density of groups based on their 
X-ray luminosities is a poor
probe of cosmological parameters (eg. Balogh \etal 1999). 
In an open Universe, most groups are predicted to form at z$\gtrsim$2-3 
(eg. Governato, Tozzi \& Cavaliere 1996). Even in
an $\Omega$=1 Universe, most groups are predicted to form at epochs
beyond the reach of current observations (eg. Eke, Cole \& Frenk 1996).

However, the thermal history of the X-ray gas in groups may be significantly affected
by non-gravitational energy input from AGN or supernovae at a level which 
is insignificant in 
the deep potentials of massive clusters.
This non-gravitational energy input can be probed via measurements of
the evolution of the X-ray properties of groups.
The X-ray luminosity of the IGM is sensitive to the 
degree and epoch of the energy injection. Energy injection inflates the 
X-ray gas, reducing its density and hence its X-ray luminosity.
Heating of the X-ray gas before it fell  into group and cluster potentials
was originally suggested (Kaiser 1991, Evrard \& Henry 1991) in order to 
break the self-similar scaling prediction
(Kaiser 1986) of strong positive evolution of the XLF; the X-ray luminosity
of groups would be decreased by a larger factor than that of clusters.
Non-gravitational energy injection also explains the shape of the 
observed L$_X$-T relation and the entropy floor of groups, as discussed 
in the introduction.   

The origin and epoch of the energy injection is unknown. Kinetic energy from AGN jets or
outflows, or supernova-driven galaxy winds, have been suggested
(eg. Valageas \& Silk 1999), either before (eg. Ponman \etal 1999), during, 
or after cluster 
collapse (eg. Lowenstein 2000). The observed lack of evolution of the 
X-ray luminosity function to z=0.5 at the low luminosities of galaxy groups 
suggests that the thermal properties of the X-ray gas has not significantly 
altered over the corresponding look-back time of $\approx$6 Gyr. Thus the 
epoch of energy injection must have been largely at z$>$0.5.

\subsection{Energy injection history based on the star formation rate}

Several authors (Menci \& Cavaliere 2000,
Wu \etal 2000, Bower \etal 2000) have made theoretical
progress by extending semi-analytic models of galaxy formation and 
evolution to include the hot, X-ray emitting gas phase predicted to be
at  the virial
temperatures of halos of galaxy mass and larger. In the models, the cooling of this gas 
via X-ray emission allows a galaxy to form. The cooling gas may be reheated by
stellar and supernova-driven winds, whose strength
depends on the star formation rate in the galaxy.
The mass of infalling gas which will form stars
is limited by the fraction of the gas which is heated, and possibly ejected, by this 
feedback mechanism. The net result is that energy is injected into the
IGM and less gas is converted into stars.
Feedback is required in the models, especially in high
redshift, low mass but dense halos, to prevent a `cooling catastrophe'
(White \& Rees 1978) and an overproduction of dwarf galaxies. 

If the Menci \& Cavaliere (2000) model parameters
controlling the 
star formation timescale and the feedback are set to produce strong 
feedback in low mass halos and match the local galaxy luminosity 
function (their model A), then the star formation rate peaks at z$\approx$1.5, and
declines at z$<$1 and z$>$2, as in the 
original Madau-Lilly plot (Madau \etal 1996). The energy input into the IGM from the 
strong feedback results in a prediction of a small amount of  evolution of the group
XLF at z=0.5, which is within the observed limits. 
Similarly, Bower \etal (2000) predict very little evolution of the 
XLF of groups if the energy input rate follows the 
star formation rate of the semi-analytical model of Cole \etal (2000).

If the Menci \& Cavaliere (2000) model feedback in low mass halos is
severely reduced (an extreme case), the star formation rate at z$>$2 becomes
constant with redshift, in agreement with extinction-corrected 
star formation histories. In this case
the hot gas is retained in shallow potentials, increasing L$_X$,
and strong positive evolution of the XLF  at group luminosities is predicted
(a factor $\approx$30 at z=1 and L$_X$=10$^{42.5}$ \ergps). At z=0.5,
the epoch observed here,
the evolution predicted is significantly less than at z=1 
(a factor of $\approx$4.5; Menci,
private communication) but is still only marginally   
consistent with the observed XLF  (and may also be inconsistent with the residual
X-ray background at 0.25 keV; Pen 1999).
Our observations thus support the need for feedback in semi-analytic models.

\subsection{Other energy injection histories}

Not all the excess energy may come from supernovae and star formation-related
activity. Several authors (eg. Valageas \& Silk 1999, Wu \etal 2000, 
Kravtsov \& Yepes 2000, Bower \etal 2000) have emphasized that the 
energy required ($\sim$0.5-3 keV per particle, depending on the epoch) 
can only be supplied by supernovae
if most of the kinetic energy of each explosion goes into heating the IGM. AGN, on the
other hand, can easily meet the energetic requirements.  

The X-ray properties of groups at high redshifts can be used to constrain the 
epoch of energy injection, irrespective of the origin of the energy.
More energy input is required (to produce the excess entropy of Ponman \etal 1999) if it 
is injected at high redshifts when collapsed systems were denser. More 
energy input inflates  the X-ray gas, decreases the central density,
and produces lower X-ray luminosities. 
If all the non-gravitational heating occurs at very high redshifts (z$>$2), 
Bower \etal (2000) predict that at z$\approx$2
groups have significantly
lower luminosities (for a given temperature) 
compared to z=0, and that the space density of low luminosity 
(L$_X$=10$^{43}$ \ergps) systems is lower than at z=0.
However, only a very small degree of evolution of the XLF is predicted 
between z=0 and z=0.5 (Bower, private communication), consistent with
the observations.
Other, preheating, models, in which the energy injection occurred at very early epochs, 
before the groups collapsed,
  similarly predict little or
no evolution of the XLF at z$\sim$0.5 and L$_X$=10$^{43}$ \ergps
(eg. Kaiser 1991, Bower 1997).

If the rate of energy injection is constant with
redshift,  
the space density of L$_X$=10$^{43}$ \ergps systems is predicted to be a factor  
$\approx$3 higher at z=0.5 than at z=0 (Bower, private communication),
just consistent with the observed XLF.

\subsection{Implications for future and current deep cluster surveys}

Whilst the results described here were based on  deep $ROSAT$ surveys, 
they represent 
the population which serendipitous Chandra, and particularly XMM-Newton, 
surveys using {\it
typical} exposures are sampling. Thus the properties of large numbers of 
very low-luminosity
groups at moderate redshifts, 
and poor clusters at high redshifts (z$\ga$1), will soon become accessible.
Measurements of 
the evolution of large-scale
structure will be possible with contiguous surveys. We have already started
to sample the large-scale  structure in this single field.
A large improvement in the statistical accuracy of the XLF will be achieved,
reaching lower luminosities than sampled here, and with 
sufficient spatial resolution
to remove the X-ray emission from AGN within the groups.  
Important additional diagnostics will be provided by deep XMM-Newton surveys,
which  will measure the evolution of the group L$_X$-T 
relation and metallicities of groups at high redshifts.
Major observational advances in the evolution of groups of galaxies are
expected.

\section{Conclusions}

We have identified groups and poor clusters of galaxies, via
their X-ray emission, at high redshifts z$\sim$0.3-0.6.
These bound systems, confirmed as such by their X-ray emission,
represent some of the lowest X-ray luminosity (and probably
lowest mass) overdensities yet found at moderate to high redshifts.
These observations support, and extend to groups of galaxies, 
the growing consensus from several deep cluster surveys that there is
little or no evolution of the cluster X-ray luminosity function 
to high redshifts (z$\sim$0.8)
at moderate X-ray luminosities. 

The X-ray luminosity evolution of low mass galaxy groups
is particularly sensitive to the thermal history of the X-ray gas,
including non-gravitational processes. Massive, luminous clusters
are better suited to deriving   
cosmological parameters.  Our results are consistent either with no 
evolution, or slight positive evolution, of 
the X-ray luminosity function at the luminosities of groups
and poor clusters, to z=0.5.
The evolution of the group X-ray luminosity function can constrain
the epoch of non-gravitational energy injection into the 
intra-group medium (due to supernova-driven winds or AGN).
The current results suggest that any such energy injection occured 
mostly at redshifts z$>$0.5. Examples of energy injection histories
consistent with our results are those which follow 
the star formation rate, at least in semi-analytical models of galaxy 
formation which include feedback, and preheating models  in which
the energy injection occured at z$\gtrsim$2. 

The identification of groups and poor clusters of galaxies at  
high redshifts also opens the way for studies of galaxy evolution in
a common, but poorly studied, environment. 
Galaxy evolution in groups is predicted to differ from
evolution in rich clusters or in the field because the low velocity
dispersion
and relatively high galaxy density favour merging, tidally-triggered
star formation and other galaxy-galaxy interactions.\\

\section{Acknowledgements}
We are grateful to Harald Ebeling 
for providing VTP flux corrections and to Nicola Menci and
Richard Bower for providing
model predictions.   The VTP source searching was initially
done for the WARPS cluster survey (Scharf \etal 1997).
This research has made use of data obtained from the 
Leicester Database and Archive Service at the Department of Physics 
and Astronomy, Leicester University. Analysis was performed on
Starlink computing facilities. LRJ acknowledges PPARC support.

\section{REFERENCES}

Allington-Smith J.R., Ellis R.S., Zirbel E.L., Oemler A., 1993, ApJ, 404, 521\\
Avni Y., Bachall J.N., 1980, ApJ, 235, 694\\
Bahcall N.A., 1981, ApJ, 247, 787\\
Balogh M.L., Babul A., Patton D.R., 1999, MNRAS, 307, 463\\
Bower R.G., Bohringer H., Briel U.G., Ellis R.S., Castander F.J., Couch W.J.,
 1994, MNRAS, 268, 345\\
Bower R.G., Hasinger G., Castander F.J., Aragon-Salamanca A., Ellis R.S.,
 Gioia I.M., Henry J.P., Burg R., Huchra J.P., Bohringer H., Briel U.G.,
 McLean B., 1996, MNRAS 281, 59\\
Bower R.G., 1997, MNRAS, 288, 355\\
Bower R.G., Benson A.J., Baugh C.M., Cole S., Frenk C.S., Lacey C.G.,
 2000, MNRAS submitted, astro-ph/0006109\\
Bryan G.L., 2000, ApJ, 544, L1 \\
Burg R., Giacconi R., Huchra J., MacKenty J., McLean B., Geller M., Hasinger G.,
 Marzke R., Schmidt M., Trumper J., 1992, A\&A 259, L9\\
Burke D.J., Collins C.A., Sharples R.M., Romer A.K., Holden B.P.,
 Nichol R.C., 1997, ApJ, 488, L83\\
Burns J.O., Ledlow M.J., Loken C., Klypin A., Voges W., Bryan G.L.,
 Norman M.L., White R.A., 1996, ApJ, 467, L49\\
Canizares C.R., Fabbiano G., Trinchieri G., 1987, ApJ 312, 503\\
Cavaliere A., Menci N., Tozzi P., 1997, ApJ, 484, L1\\ 
Cavaliere A., Menci N., Tozzi P., 1999, MNRAS, 308, 599\\
Cole S.,  Lacey, C.G., Baugh C.M., Frenk C.S., 2000, MNRAS, 319, 168\\
Collins C.A. \& Mann R.G. 1998, MNRAS, 297, 128.\\
De Grandi S., Guzzo L., Bohringer H., Molendi S., Chincarini G.,
 Collins C., Cruddace R., Neumann D., Schindler S., Schuecker P., Voges W.,
 1999, ApJ, 513, L17\\ 
Ebeling H. \& Wiedenmann G., 1993, Phys. Rev. 47, 704\\
Ebeling H., Edge A., Fabian A.C., Allen S.W., Crawford C.S., Bohringer H.
 1997, ApJ, 479, L101\\
Eke V.R, Cole S., Frenk C.S., 1996, MNRAS, 282, 263\\
Evrard A.E. \& Henry J.P., 1991, ApJ, 383, 95\\
Fairley B.W., Jones L. R., Scharf C., Ebeling H.,
 Perlman E., Horner D., Wegner G., Malkan M., 2000, MNRAS 315, 669\\
Fukugita M., Hogan C.J., Peebles P.J.E., 1998, ApJ, 503, 518\\
Governato F., Tozzi P., Cavaliere A., 1996, ApJ, 458, 18\\
Hasinger G., Burg R., Giacconi R., Schmidt M., Trumper J., Zamorani G., 1998a,
 A\&A, 329, 482\\
Hasinger G., Giacconi R., Gunn J.E., Lehmann I., Schmidt M., Schneider D.P.,
 Trumper J., Wambsganss J., Woods D., Zamorani G., 1998b, A\&A 340, L27\\
Henry J.P., Gioia I.M., Huchra J.P., Burg R., McLean B., Bohringer H., 
 Bower R.G., Briel U.G., Voges W., MacGillivray H., Cruddace R.G., 1995,
 ApJ, 449, 422\\
Jones L.R., Scharf C.A., Ebeling H., Perlman E., Wegner G.,
 Malkan M., Horner, D., 1998, ApJ, 495, 100\\
Jones L.R., Ebeling H., Scharf C.A., Perlman E., Horner, D., Fairley B.,
 Wegner G., Malkan M. 2000a, in Durret F. \& Gerbal D., eds., Constructing the Universe
 with clusters of galaxies. Institute 
 d'Astrophysique de Paris, Paris\\
Jones L.R., Ponman T.J., Forbes D.A., 2000b, MNRAS, 312, 139\\   
Kaiser N., 1986, MNRAS, 222, 323\\
Kaiser N., 1991, MNRAS, 383, 104\\
Kravtsov A.V. \& Yepes G., 2000, MNRAS, 318, 227\\
Lehmann I., Hasinger G., Schmidt M., Gunn J.E., Schneider D.P., Giacconi R.,
 McCaughrean M., Trumper J., Zamorani G., 2000, A\&A, 354, 35\\
Lloyd-Davies, E.J., Ponman, T.J., Cannon, D.B., 2000, MNRAS 315, 689\\
Lowenstein, M., 2000, ApJ, 532, 17\\
Madau P.,  Ferguson H. C., Dickinson M. E.,
 Giavalisco M., Steidel C. C., Fruchter A., 1996, MNRAS, 283, 1388\\
Menci N. \& Cavaliere A., 2000, MNRAS, 311, 50\\
McHardy I.M., Jones L.R. et al, 1998, MNRAS, 295, 641 (M98)\\
Mulchaey J.S., Davis D.S., Mushotzky R.F., Burnstein D., 1996,
ApJ, 456, 80\\
Mushotzky R.F., Cowie L.L., Barger A.J.,
 Arnaud K.A.,  2000, Nature, 404 459 \\
Newsam A.M., McHardy I.M., Jones L.R., Mason K.O., 1997,
MNRAS, 292, 378\\
Pen U-L., 1999, ApJ, 510, L1\\
Ponman T.J., Bourner, P.D.J., Ebeling, H., Bohringer H., 1996, MNRAS,
 283, 690\\
Ponman T.J., Cannon D.B., Navarro J.F., 1999, Nat, 397, 135\\
Postman M., Lubin L.M., Gunn J.E., Oke J.B., Hoessel J.G., 
 Schneider D.P., Christensen J.A., 1996, AJ 111, 615\\
Romer A.K., Nichol R. C., Holden B. P., Ulmer M. P.,
 Pildis R. A., Merrelli A. J., Adami C., Burke D. J.,
 Collins C. A., Metevier A. J., Kron R. G., Commons K., 2000, ApJS, 126, 209\\
Rosati P., Della Ceca R., Norman C., Giacconi R., 1998a, ApJ, 492, L21\\
Rosati P. 1998b, in proceedings of The Young Universe: Galaxy Formation 
 and Evolution at Intermediate and High Redshift, ASP conference
 series, ed D'Odorico S., Fontana A., Giallongo E., 146, 476\\
Scharf C., Jones L.R., Ebeling H., Perlman E., Malkan M. \&
 Wegner G., 1997, ApJ, 477, 79\\
Schmidt M., Hasinger G., Gunn J., Schneider, D.,
 Burg R., Giacconi, R., Lehmann I., MacKenty J.,
 Trumper J., Zamorani G., 1998, A\&A, 329, 495\\
Snowden S.L., Plucinsky P.P., Briel U., Hasinger G., Pfeffermann E.,
 1992, ApJ, 393, 819\\
Tozzi P. \& Norman C., 2001, ApJ, 546, 63\\
Tully R.B., 1987, ApJ, 321, 280\\
Valageas P. \& Silk J., 1999, A\&A, 350, 725\\ 
Vikhlinin A., McNamara B.R., Forman W., Jones C., Quintana H., Hornstrup
 A., 1998, ApJ, 502, 558\\
White D.A., Jones C., Forman W., 1997, MNRAS, 292, 419\\
White S.D.M. \& Rees M., 1978, MNRAS, 183, 341\\ 
Wu K.K.S., Fabian A.C., Nulsen P.E.J., 1998, MNRAS, 301, L20\\
Wu K.K.S., Fabian A.C., Nulsen P.E.J., 2000, MNRAS, 318, 889\\
Zamorani G., Mignoli M., Hasinger G., Burg R., Giacconi R., Schmidt M., 
 Trumper J., Ciliegi P., Gruppioni C., Marano B., 1999, A\&A, 346, 731\\

\noindent

\section{APPENDIX A: OTHER CANDIDATE GROUPS AND CLUSTERS}

In this appendix we give descriptions of the additional candidate 
groups and clusters.

\begin{table*}
\begin{minipage}{175mm}
\caption {Properties of the additional candidate groups and clusters}
\tiny
 \begin{tabular}{lllllllllllll} \hline
ID & \multicolumn{1}{c}{PSPC (0.5-2 keV)}& \multicolumn{2}{l}{PSPC
position (J2000)} &  
Offset & Id & Hardness & HRI & R & n$_z$  &  z & $L_X^{0.5-2}$ (10$^{42}$ & PSPC  \\
 & flux x10$^{-15}$& RA &Dec & 
(\arcsec) & class & ratio& detect & mag & &   & erg s$^{-1}$) & extended?\\
  & & & & & & & & & & & &   \\
  & [a] & [b]&[c]&[d]&[e]&[f]&[g]&[h]&[i]&[j]&[k]&[m] \\
  & & & & & & & & & & & &   \\

5b &$\la$4.9 &13 34 40.2 & 37 55 50 &  - &   & - & No &  & 0 & - & & \\ 
49 & 10.4  &13 34 46.97 & 37 47 48.4$^{\dag}$ & 0.6& * & $>$1.10&pt&21.1&1&0.709&28& \\  
51 & 11.4  &13 33 59.88 & 37 49 11.0$^{\dag}$& 0.7 & (*)& 0.54$\pm$0.10&pt&17.8&2&0.257&3.7&?\\
62 &9.3   & 13 35 09.32 & 37 48 21.7 & 9.3 & * & 0.62$\pm$0.14& No &
18.6 & 1 & 0.251 &2.9 & \\
77 &7.1   & 13 35 32.63 & 37 45 49.0 & 15.4 & (*)& 0.47$\pm$0.12 & No &
18.1 & 2 & 0.307 &3.5 & \\
99$^{\ddag}$ & 5.2 &13 35 3.81 & 37 44 18.3$^{\dag}$&  2 & (*) & $>$0.52&Yes&23&0&$\ga$1.3&$\ga$52 & \\
105$^{\ddag}$& 4.2 &13 34 19.00& 37 40 18.6 & 3.5& (*)&0.33$\pm$0.09&No&18.6&0&-&-& \\
112& 5.1 &13 34 35.85 & 37 54 22.7 & 8 &    & $>$0.43&No& 23.1&0&$\ga$1.3&$\ga$51 & \\
\hline
 \end{tabular}
 \normalsize

$^{\ddag}$ Not in complete sample area (see Fig 2 of M98)\\
$^{\dag}$ HRI position
 \end{minipage}
 \end{table*}

~\\
\noindent{\bf 5.} M98 noted that this source is in a confused area. The
PSPC X-ray source consists of a dominant point-like source
with a fainter, possibly extended, component to the SW. There are several pieces
of evidence suggesting that the dominant source is not a cluster.
The PSPC hardness ratio
(0.37$\pm$0.03) is inconsistent with the hardness ratios of the 
confirmed group and clusters, and with that predicted for thermal 
X-ray spectra (see M98). The HRI data are consistent with a point-like
source, and finally the PSPC count rates
in both the hard and soft bands vary significantly (and consistently)
between the two PSPC observation epochs, separated by 2 years. 
After measuring and subtracting the point-source flux within an aperture 
of radius 30 arcsec (containing 90\% of the  1 keV flux), the 
remaining total extended PSPC flux 
is $\la$4.9x10$^{-15}$ \ecs, below the flux limit considered here.
As noted by M98, this source
(which we label 5b) is coincident with an excess of faint galaxies,
the redshifts of which are unknown. A nearby stellar object is
also a possible counterpart to source 5b.

\noindent{\bf 49.}
An absorption-line galaxy at a redshift of z=0.709 lies only 0.6 arcsec
from the HRI position, which confirms the PSPC counterpart. The optical
spectrum of the galaxy is of low signal-noise.
No excess of galaxies is visible to a limit in R
$\approx$3.5 mag fainter than this galaxy. The mere fact of detection in the 
HRI suggests that the X-ray emission does not originate in an extended
intra-cluster medium, but in a point source 
associated with the galaxy. The X-ray luminosity
(3x10$^{43}$ \ergps) is, however, much higher than expected for a normal
galaxy, and this source may be similar to the sources  found in deep Chandra surveys 
which have optically unremarkable late-type galaxy counterparts 
(eg. Mushotzky \etal 2000). 
It is unlikely to originate in a cluster. 

\noindent{\bf 51.}
The X-ray source has at least 2 components. An HRI detection of a point source
coincident with an absorption line galaxy at z=0.257 (confirming the
PSPC position)  suggests that the 
tentative identification of M98 with a narrow emission line galaxy was 
incorrect. As noted by M98, the X-ray data are consistent with two point 
sources, although some contribution from an extended intra-group medium is possible.
There are several R=18-19 mag galaxies nearby, but none of the four 
measured redshifts  
falls within 1000 km s$^{-1}$ of any of the others, suggesting that the 
excess of galaxies is partly a projection effect. The HRI detection
(with a flux consistent with the PSPC flux) suggests that most of the 
X-ray emission originates in
the z=0.257 absorption line galaxy rather than in an intra-group medium.
The galaxy is luminous optically (M$_R$=-23.5) and in X-rays (L$_X$=4x10$^{42}$
\ergps) and has log($L_X^{bol}/L_B$)$\approx$-1.8, 
higher than that of any of the 81 early type galaxies studied by  Canizares,
Fabbiano \& Trinchieri (1987). It may contain an AGN, like the
early type galaxy counterparts to
sources detected in deep Chandra surveys, although no optical
emission lines are observed.

\noindent{\bf 62.}
A luminous (M$_R$=-22.7) galaxy at a redshift
of z=0.251 lies 9 arcsec from the PSPC position and there is a small 
excess of fainter
galaxies (R$>$19) nearby. The X-ray source is unresolved by the PSPC, but 
is undetected
by the HRI, suggesting that it is either a low luminosity ($L_X$=2.9x10$^{42}$ \ergps),
compact extended source (of extent
$\approx$10-20 arcsec or 50-100 kpc at z=0.251), or a variable point source.  
There are no optical objects of stellar appearance with R$<$24.5 in the error
circle, which could be QSOs, but rather several faint (R$\sim$22) galaxies.
The source could be a poor group dominated by the luminous galaxy (with the 
lowest X-ray luminosity of any system in the group/cluster sample),
but the positional offset of the luminous galaxy makes this uncertain.

\noindent{\bf 77.}
A very luminous elliptical galaxy of $M_R\approx$-23.7 ($\approx$3L$^*$)
with an absorption line spectrum at a redshift
of z=0.307 is the probable counterpart to the X-ray source. A second, nearby,
fainter galaxy has the same  redshift. The lack of an obvious excess of galaxies to
a limit 6 magnitudes fainter than that of the bright galaxy
and the low X-ray luminosity (3.4x10$^{42}$ \ergps) suggest that a large fraction
of the X-ray emission may originate in an individual elliptical galaxy rather 
than in an intra-group medium, although the value of log($L_X^{bol}/L_B$)$\approx$-1.9
is higher than that of any of the early type galaxies of
Canizares \etal (1987). The source may be a fossil group of the type studied
by Jones et al. (2000b).
The offset between the X-ray centroid and the brightest galaxy is 
large (15 arcsec), but this source is the furthest off-axis in the PSPC
of any in the sample. At this off-axis angle (14 arcmin), the PSPC 1 keV PSF is
almost double the size of the on-axis value, so larger position errors
are expected. Additional position errors may have occurred because source 98 
(of M98) is only 50 arcsec away (ie the FWHM of the off-axis PSF).
Alternatively, the counterpart may be a variable AGN, 
undetected in the HRI and with R$>$24.5 mag at the epoch of the optical imaging.

\noindent{\bf 99.}
Although this source is outside the complete survey area,
it is potentially of interest. It is listed by M98 as a blank field (ie
no counterpart with R$<$23 mag). A possible extremely distant cluster
of galaxies lies 20 arcsec south of the PSPC position. The likely brightest
cluster galaxy (BCG) has R$\ga$23 mag, giving an estimated redshift z$\ga$1.3,
where the redshift has been estimated from the BCG magnitude by
extrapolating  the relation of Vikhlinin \etal (1998).
However, the detection in the HRI data of a X-ray source coincident with a
R=23 mag galaxy 9 arcsec south of the PSPC position suggests that this
galaxy is the counterpart, and not the intra-cluster medium of the 
possible distant cluster.

\noindent{\bf 105.}
This source is also outside the complete survey area but we list it here
because 
a compact group of three galaxies of R$\approx$18 mag lie within the
PSPC error box. The redshifts are unknown, but based on the BCG
magnitude probably lie in the range z=0.25-0.3. 

\noindent{\bf 112.}
This source is potentially an extremely distant cluster. It is listed by
M98 as a blank field. A galaxy of R=23.1 mag surrounded by several
fainter galaxies lies within
the PSPC error circle. Near infra-red K band imaging (Newsam \etal 1997) 
shows that many of the galaxies of very red; the brightest has
R-K=4.2$\pm$0.4 and R-I=1.6 (to be compared with R-K=2.6 and R-I=0.6
for zero redshift ellipticals). If this galaxy  is the BCG of a cluster,
then the redshift estimated from its R band magnitude is z$\ga$1.3,
and from its K=18.9 magnitude, involving a less uncertain 
extrapolation, z$\approx$1.5
(using fig 2 of Collins \& Mann 1998).
The X-ray luminosity of the cluster, even at this high redshift, is
$\sim$10$^{44}$ \ergps, so it would be a typical, sub-L$^*$ cluster rather
than a luminous, rich cluster.

\end{document}